\documentclass[aps,prd,nofootinbib,twocolumn,floatfix,letterpaper,superscriptaddress,showpacs]{revtex4}
\usepackage{graphicx}\usepackage{natbib}\usepackage{amsmath}
\pdfoutput=1
\usepackage{times}

\newcommand{\aj}{Astronom.~J.\ }
\newcommand{\apjs}{Astrophys.~J.~Suppl.\ }
\newcommand{\mnras}{Mon.~Not.~Roy.~Astron.~Soc.\ }
\newcommand{\aap}{Astron.~Astrophys.\ }

\begin{document}

\title{Problems and Prospects from a Flood of Extragalactic TeV Neutrinos in IceCube}

\author{Matthew D. Kistler}
\affiliation{Kavli Institute for Particle Astrophysics and Cosmology, Stanford University, SLAC National Accelerator Laboratory, Menlo Park, CA 94025}

\date{November 3, 2015}

\begin{abstract}
The steep spectrum of neutrinos measured by IceCube extending from $>\,$1~PeV down to $\lesssim\,$10~TeV has an energy flux now encroaching on the {\it Fermi} isotropic GeV background.  We examine several implications starting from source energetics requirements for neutrino production.  We show how the environment of extragalactic nuclei can extinguish $\sim\! 10 \!-\! 100\,$TeV gamma rays and convert their energy to X-rays for plausible conditions of infrared luminosity and magnetic field, so that the {\it Fermi} background is not overwhelmed by cascades.   We address a variety of scenarios, such as for acceleration by supermassive black holes and hadronic scenarios, and observations that may help elucidate the neutrinos' shadowy origins.
\end{abstract}

\pacs{98.70.-f, 98.70.Rz, 98.70.Sa, 95.85.Ry}
\maketitle

\section{Introduction}
While it was appreciated early that cosmic-ray interactions produce secondary particles from meson decays (e.g., \cite{Feenberg1948,Fermi1949,Richtmyer1949}), neutrinos remained scarcely mentioned (cf., \cite{Burbidge1956}) until the discovery of reactor antineutrinos \cite{Cowan1956}.  A long history of flux predictions and proposed means of detection that soon began (e.g., \cite{Greisen:1960wc,Reines:1960we,Markov1961,Bahcall1964,Berezinsky:1975zz,Eichler:1978zn,Stecker:1978ah,Roberts:1992re,Halzen:1988wr,Barwick:1991ur,Gaisser:1994yf,Waxman:1998yy,Learned:2000sw,Halzen:2002pg,Becker:2007sv}) has been capped by IceCube \cite{Ahrens:2002dv} actually detecting a high-energy astrophysical neutrino flux, presenting new challenges and opportunities.  The accumulation of data has progessed from but two $E_\nu \!\simeq\! 1$~PeV events at the start \cite{Aartsen2013}, to a $\sim\! E_\nu^{-2}$ flux \cite{Aartsen2013b}, getting softer \cite{Aartsen2014,Aartsen2015} to $\sim\! E_\nu^{-2.7}$ \cite{Niederhausen2015} at present.
The net result is a $\sim\! 10 \!-\! 100$~TeV neutrino flux that is much larger than previously thought \cite{Aartsen2015,Niederhausen2015}.

\begin{figure*}[t!]\vspace*{-0.1cm}
\includegraphics[width=1.9\columnwidth,clip=true]{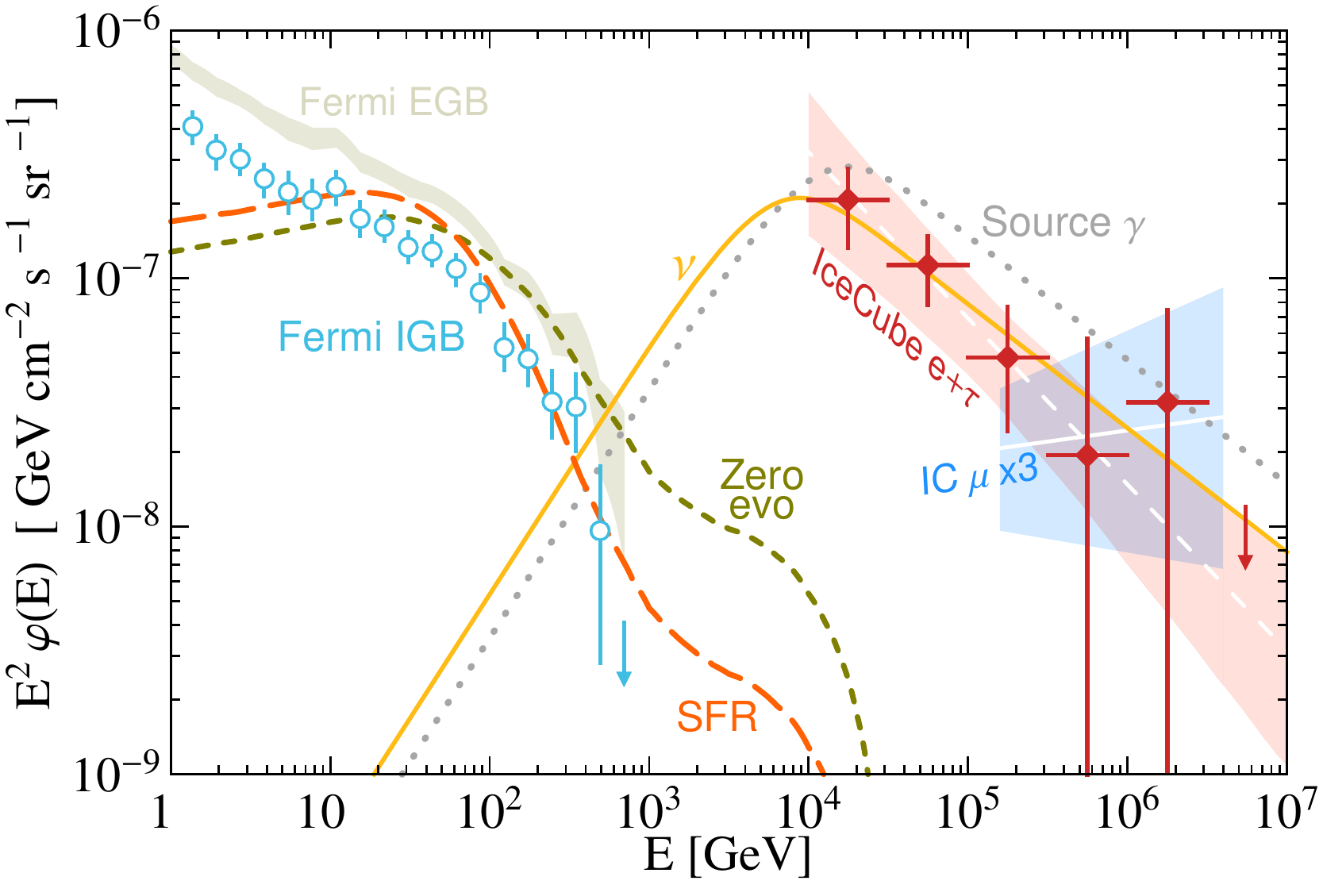}
\vspace*{-0.3cm}
\caption{Minimal models for an extragalactic explanation of the $\gtrsim\! 10$~TeV IceCube neutrino data ({\it solid line}) with the corresponding gamma-ray flux at production if attenuation on source backgrounds and extragalactic background light (EBL) were both neglected ({\it dotted line}).  Including EBL cascades with either zero cosmic source evolution ({\it short dashed line}) or star formation rate evolution ({\it long dashed line}) overproduces the IGB.  We show the {\it Fermi} EGB ({\it band}) that includes  sources \cite{Ackermann2015}, which, if saturated, would imply neutrino correlations with {\it Fermi} blazars.
\label{casca}}
\end{figure*}

On the extragalactic front, it is remarkable that the IceCube TeV flux is of the same order as the {\it Fermi} isotropic gamma-ray background (IGB) \cite{Ackermann2015} (see Fig.~\ref{casca}).  Indeed, the neutrinos most likely result from the decays of pions made in $p \gamma$ or $pp$ scattering, which must also result in gamma rays.  Comparing measurements of the ultrahigh-energy cosmic-ray flux to the IceCube data imply that the cosmic-ray sources may be optically thin to photopion losses (\cite{Kistler2014}; see also \cite{Ahlers2009}).  The photon background required for this process may yet imply that $\gamma \gamma \!\rightarrow\! e^+ e^-$ extinction within the neutrino production sites can be important (e.g., \cite{Mannheim2001,Kistler2014,Murase2015b}), although extrapolation to the $\sim\!10-100$~TeV range is not necessarily direct.

If these gamma rays can escape their source, they run the risk of initiating cascades on intergalactic starlight leading to a $\lesssim\!100$~GeV flux in excess of the {\it Fermi} IGB, acutely so if their spectrum extends softly to GeV energies \cite{Murase2013}.  This presents all sorts of difficulties, particularly to an extragalactic neutrino interpretation and connections such as to ultrahigh-energy cosmic rays (e.g., \cite{Kistler2014,Baerwald:2013pu,Anchordoqui2014,Baerwald2015}).

Our purpose is to address some of what can be determined from the IceCube data and conditions that must be satisfied.  In particular, we consider the requirements placed by the IceCube flux on energetics and whether prospective source classes that may produce neutrinos in this range, such as various types of AGN or starburst galaxies, can handle the load.

Drawing from experience with our Galactic Center \cite{Kistler2015}, we discuss the efficacy of infrared backgrounds for $\gamma \gamma \!\rightarrow\! e^+ e^-$ near the centers of other galaxies, finding that plausible infrared luminosities can quench the TeV gamma rays produced along with neutrinos, so that even if the EGB is near-saturated by blazars \cite{Ajello2015,Ackermann2015b,Bechtol2015} neutrino production can still be accommodated.  Provided that magnetic fields are sufficiently strong, the result is a hard X-ray, rather than GeV, flux that may be utilized as a signal for the neutrino sources.

\section{River of Neutrinos}
We begin by showing in Fig.~\ref{casca} the most recent presented IceCube spectral data.  These include a search for $\nu_e + \nu_\tau$ showers that has uncovering more $\gtrsim\! 10$~TeV events a led to a soft $\sim\! E_\nu^{-2.67}$ spectrum \cite{Niederhausen2015}.  Muon measurements imply a hard $\nu_\mu$ spectrum \cite{Aartsen2015b}, $\sim\! E_\nu^{-1.9}$ \cite{Radel2015}, although these only address higher energies thus far.  While it is possible that peculiar flavor behavior can occur in the muon range \cite{Kistler2014,Winter:2013cla,Winter:2014pya}, we will not address this further here (or fluxes at much higher energies \cite{Beresinsky:1969qj,Stecker:1978ah,Hill:1983mk,Yoshida:1993pt,Engel:2001hd}) and stick to the TeV regime assuming equality between flavors and $\nu/\bar{\nu}$.

The sky distribution of IceCube neutrinos is most easily interpreted as being extragalactic in origin \cite{Aartsen2013,Aartsen2013b,Aartsen2014}, though we are in the unusual situation of only knowing a neutrino flux.  We begin by constructing a heuristic model.  To do so we make an assumption on the spectrum and relative number of charged and neutral pions produced.  It is easy to see by eye in Fig.~\ref{casca} that extending even a flat ($E_\gamma^{-2}$) gamma-ray spectrum, as might be expected from $pp$ scattering, back to GeV energies from $\sim\!10$~TeV will result in conflict with the {\it Fermi} IGB.  Hence, we use a break at $\sim\! 10$~TeV (as does \cite{Murase2015b}).  This could be indicative of photopion production near $p\gamma$ threshold, with the break due to a lack of higher-energy target photons for lower-energy protons.

Photopion neutrinos with $E_\nu\!\sim\!10$~TeV would result from $E_p\!\gtrsim\! 200$~TeV protons, the $p\gamma$ threshold (neglecting possible relativistic beaming effects for now) requiring $\lesssim\,$keV X-rays for targets.  Thus, a turndown due to a lower number density of more energetic target photons within the source appears plausible.  We assume $p \gamma$ production near threshold, resulting in mostly $\pi^+$ (which could also ease tension with a lack of Glashow resonance events in IceCube \cite{Kistler2014}).  Assuming equal numbers of $\pi^+$, $\pi^-$, and $\pi^0$ results in half the gamma-ray flux (since only the total charged pion production is fixed by the neutrino flux measurement).

We use a smoothly-broken power law to describe the source
\begin{equation}
      \frac{dN_\nu}{dE_\nu}  =   f_\nu
       \left[\left(E/E_b\right)^{\alpha \eta} + \left(E/E_b\right)^{\beta \eta} \right]^{1/\eta} \,,
\label{specfit}
\end{equation}
obtaining fluxes at Earth, $\varphi_\nu(E_\nu)$, by integrating this as
\begin{equation}
  \varphi_\nu(E_\nu) = \frac{c}{4 \pi } \int_0^{z_{max}} \frac{dN_{\nu}}{dE_{\nu}^\prime}  \frac{dE_{\nu}^\prime}{dE_{\nu}}\, \frac{\mathcal{W}(z)}{dz/dt} \,dz \,,
\label{f2}
\end{equation}
where ${dz}/{dt} \!=\! H_0\, (1 \!+\! z) [\Omega_m (1 \!+\! z)^3 \!+\! \Omega_\Lambda ]^{1/2}$, ($\Omega_m \!=\! 0.3$, $\Omega_{\Lambda} \!=\!0.7$, and ${H}_{0} \!=\! 70\,$km/s/Mpc), and $dE_\nu^\prime/dE_\nu \!=\! (1+z)$.

Source evolution with redshift, $\mathcal{W}(z)$, does not greatly affect the arriving neutrino spectral shape.  To obtain the neutrino flux in Fig.~\ref{casca} with zero evolution ($\mathcal{W}(z) \!=\! 1$), we use slopes $\alpha \!=\! -1$ and $\beta \!=\! -2.5$, broken at $E_b \!=\! 10$~TeV, and $\eta \!=\! -2$ to break smoothly.  To obtain an equivalent $\varphi_\nu(E_\nu)$ using cosmic star formation rate evolution (SFR) \cite{Hopkins2006,Yuksel2008,Kistler2009b} requires lowering the source normalization by a factor of $\sim\!4$ and a slight shift of $E_b$ to $\sim\! 13$~TeV.

We examine the required cosmic neutrino emissivity to arrive at the flux in Fig.~\ref{casca} for some guidance.  Assuming zero source evolution, we find $\mathcal{E}_\nu \!\approx\! 7 \!\times\! 10^{37}\,$erg~s$^{-1}\,{\rm Mpc}^{-3}$, while SFR evolution reduces this by a factor of $\sim\! 4$ at $z \!=\! 0$.  As we will see, these are substantial and indicate at least one of the two favorite high-power sources: AGN or supernovae.

TeV gamma rays are rather different, considering that they can be attenuated by $\gamma \gamma \!\rightarrow\! e^+ e^-$ on the CMB or EBL (intergalactic starlight or infrared photons) even if they can escape their source.  We assume free escape here only for illustration of the present difficulty, assuming the same spectral slopes in Eq.~(\ref{specfit}), translating using $E_\gamma \!\approx\! 2 E_\nu$ with two gamma rays produced for every three neutrinos.

In Fig.~\ref{casca}, we show the resulting diffuse gamma-ray flux from $e^\pm$ cascades on the EBL from this input spectrum obtained using ELMAG (\cite{Kachelriess2012}; v.~2.02 with EBL of \cite{Kneiske:2003tx}) for both zero and SFR evolution scenarios.  We see that these already saturate or exceed the {\it Fermi} IGB at various energies.  Stronger evolution places more production at higher $z$, decreasing the low-$z$ burden and causing the sharper decline at $E_\gamma \!\gtrsim\! 100$~GeV, with a greater accumulation at $\lesssim\! 100$~GeV.

This is clearly an issue \cite{Murase2013}, since the IGB does not include any contribute from unresolved sources, which should be present at some level.  For instance, much of the total {\it Fermi} extragalactic background (EGB) that includes extragalactic sources \cite{Ackermann2015} is due to blazars, while \cite{Ajello2015} argues that extrapolating the blazar luminosity function to below the {\it Fermi} source threshold contributes a sizable fraction of the IGB.

Now, in principle, even the EGB could be saturated in our case.  However, this would imply a large contribution from the neutrino production mechanism to the gamma-ray flux of presently known sources, which become fewer in number as the highest {\it Fermi} energy bin is approached \cite{Ackermann2015}.  Thus, some level of correlation of IceCube neutrinos with {\it Fermi} sources, which are mostly blazars \cite{Acero2015}, would be expected.  This is not clearly evident at present.  Comparing IceCube track event positions \cite{Aartsen2013b,Aartsen2014,Kopper2015} to Second Hard Fermi-LAT catalog sources \cite{Ackermann:2015uya}, we find no coincidences within the median angular errors of the tracks.

\section{Quenching Gamma rays}
Of course, these neutrinos and gamma rays have to be produced somewhere, and that somewhere must contain some local background photon, particle, and magnetic field density.  The former is at least required for $p \gamma$ interactions, the latter may be connected to proton acceleration.  It has long been understood that the source photon background can also be applied to $\gamma \gamma$ interactions (e.g., \cite{Mannheim2001}), thus relevant in the IceCube era for multi-messenger connections to $p \gamma$ sources \cite{Kistler2014,Winter:2013cla}.  A soft X-ray background for $\sim\! 10$~TeV neutrinos suggests opaqueness to $\gtrsim\,$GeV gamma rays in particular \cite{Mannheim2001,Murase2015b}.  Beyond this requires extrapolation of the local photon field within the neutrino production site, unknown of course at this point, to higher and lower photon energies.

We consider here an environment that such sources may be embedded within, which can lead to attenuation largely independent of what is occurring within the acceleration region.  Of course, every source is peculiar in its own way, and a first principles model of the energy/angle dependent photon field would thus be source dependent.  We use our own Galactic Center (GC) as a starting point, with the phenomenological GC background obtained in \cite{Kistler2015} from recent data, including {\it Herschel} \cite{Etxaluze2011,Goicoechea2013} and {\it SOFIA} \cite{Lau2013}.  These IR data are consistent a fraction of the incident UV flux from a $T \!\approx\! 35000\,$K,  $L_{35000} \!\approx\! 2\!\times\!10^7\,L_\odot$ cluster of massive stars at the GC driving emission down into the far infrared (FIR) by dust reprocessing in the circumnuclear disk (e.g., \cite{Davidson1992,Krabbe1995,Genzel2010}).

\begin{figure}[t!]
\hspace*{-0.3cm}
\includegraphics[width=1.05\columnwidth,clip=true]{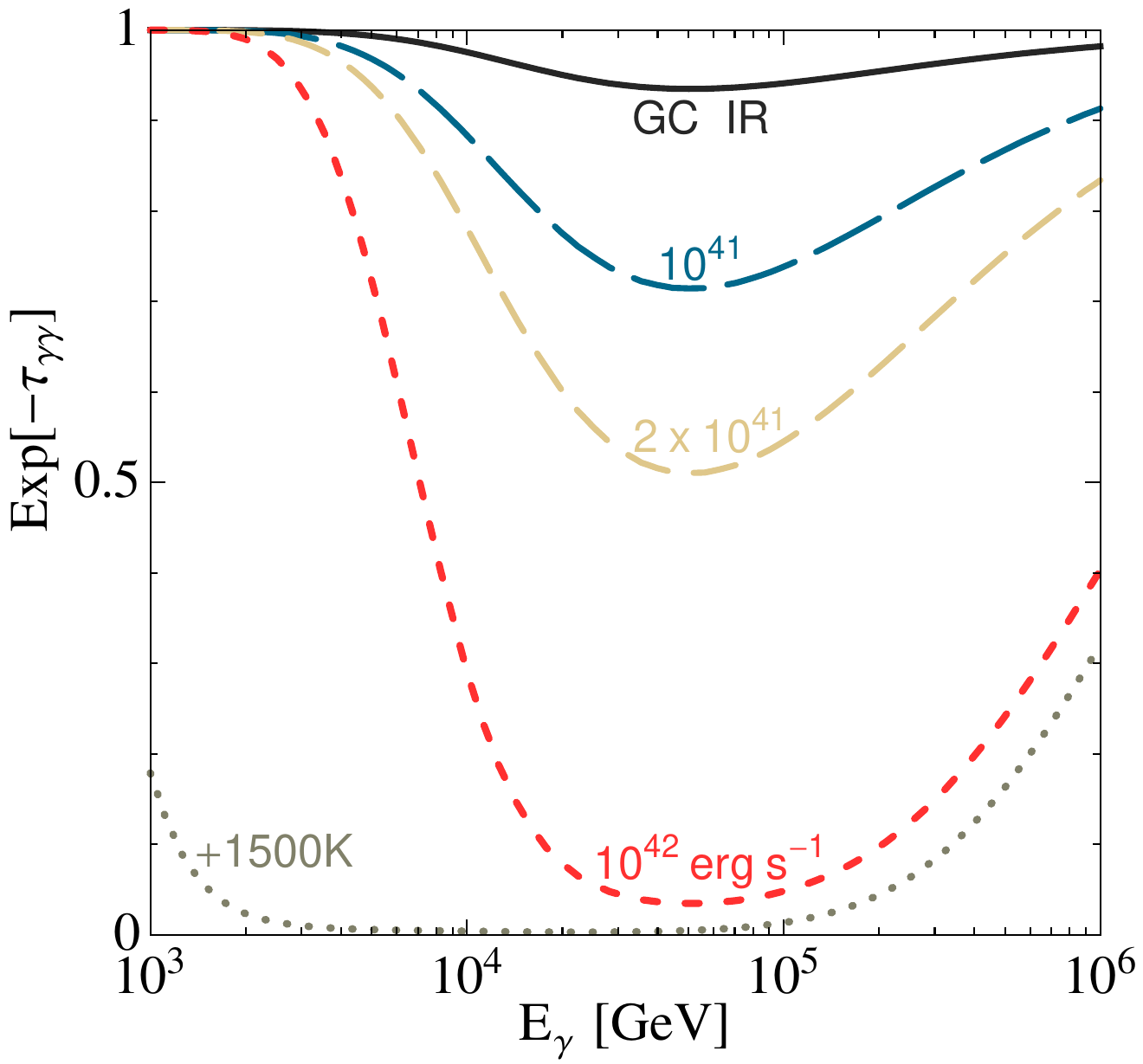}
\vspace*{-0.6cm}
\caption{Attenuation of gamma rays from our Galactic Center due to the infrared background components with $L_{\rm IR} \!\sim\! 2 \!\times\! 10^{40}\,$erg~s$^{-1}$ from \cite{Kistler2015} ({\it solid line}).  We show the effect of scaling the luminosities by a factor of 5, 10, or 50 ({\it dashed lines}), as may be more indicative of galaxies with more active inner parsec regions, and adding to the last a $\sim\! 0.5\,$pc scale $10^{43}\,$erg~s$^{-1}$ 1500~K component ({\it dotted line}) characteristic of AGN torii.
\label{ohigh}}
\end{figure}

This emission arises from $\sim\,$pc scales, so any interference with an accelerator working on much smaller scales would be minimal (e.g., negligible additional photopion cooling is introduced), while IR photon backgrounds remain rather effective at $\gamma \gamma \!\rightarrow\! e^+ e^-$ for $\sim\!10-100$~TeV gamma rays.  While the gamma-ray opacity in our GC (Fig.~\ref{ohigh}) is not overwhelming at present \cite{Kistler2015}, the IR flux could have been far larger than the current $L_{\rm IR} \!\sim\! 5 \!\times\! 10^6\, L_\odot \!\sim\! 2 \!\times\! 10^{40}\,$erg~s$^{-1}$ in the past or in more active extragalactic central parsec regions.

Fig.~\ref{ohigh} demonstrates that a modestly larger IR background compared to the GC inner parsec leads to a considerable $\gamma \gamma \rightarrow e^+ e^-$ optical depth in the energy range of interest for IceCube (for a larger region $\tau_{\gamma \gamma} \!\sim\! 1\,$pc$/R$ for the same luminosity).  We consider a scenario in which neutrino production is occurring in similar environments and $\gtrsim\! 10$~TeV gamma rays are attenuated into $e^\pm$ pairs that then radiate their energy via synchrotron or inverse Compton (IC) processes.

We use the $10^{42}\,$erg~s$^{-1}$ opacity, which itself is not remarkable (we will discuss comparisons to AGN luminosities later).  Note we need account for interactions occurring in the source frame, since we are observing a redshifted neutrino spectrum and the IR pair opacity can have a strong energy dependence, which in this instance ends up in more TeV gamma-ray attenuation.  We use the methods for electron spectral evaluation from \cite{Kistler2015}, considering equilibrium conditions and taking $E_e \!=\! E_\gamma / 2$ for $e^\pm$ pairs made by attenuated gamma rays, assuming that the electron population has a locally isotropic velocity distribution and that relativistic beaming is not relevant.

\begin{figure*}[t!]\vspace*{-0.2cm}
\includegraphics[width=2.05\columnwidth,clip=true]{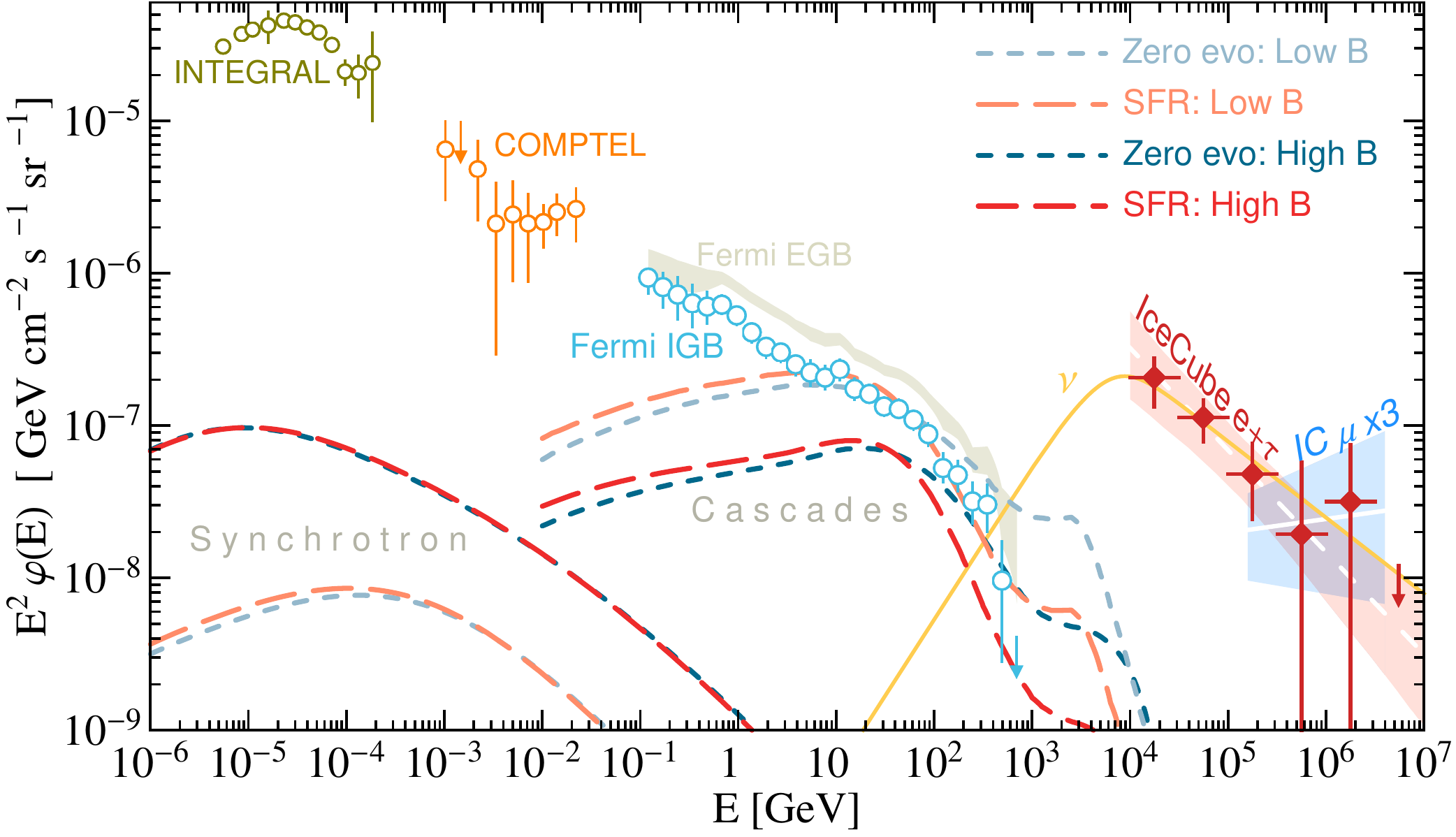}
\vspace*{-0.3cm}
\caption{Neutrino spectrum and data from Fig.~\ref{casca}, along with extragalactic {\it INTEGRAL} \cite{Churazov2007} and {\it COMPTEL} \cite{Weidenspointner2000} fluxes.
We add models incorporating source extinction via the $10^{42}\,$erg~s$^{-1}$ central parsec model using zero ({\it short dashed}) or star formation rate ({\it long dashed}) evolution, with either higher ({\it dark lines}) or lower ({\it light lines}) source magnetic field, resulting in residual EBL cascades and synchrotron radiation.
The combination of higher $B$ and SFR evolution ({\it dark, long dashed}) falls below all data (though could be further suppressed; \S\ref{whence}).
\label{stream}}
\end{figure*}

For the magnetic field, we assume only that the typical strength $B$ is within the range discussed in \cite{Kistler2015b} from observations of the GC magnetar SGR J1745--29 \cite{Eatough2013}.  We use a ``High~B'' case with $B \!=\! 3\,$mG and ``Low~B'' with $B \!=\! 0.1\,$mG as examples.  One could include secondary $e^\pm$ from pion decays in calculating synchrotron, although these should travel a much shorter distance than gamma rays or neutrinos, so are more adequately handled in descriptions of the neutrino production region, which likely has much larger $B$.  Considering the characteristic energy of synchrotron emission,
\begin{equation}
       E_\gamma \sim 20\, \left(\frac{E_e}{20\,{\rm TeV}}\right)^{\!2} \left(\frac{B}{{\rm mG}}\right) \, {\rm keV}  \notag
    \,,
\label{Echar}
\end{equation}
we see that hard X-rays can be the main end product of pair cooling, provided the magnetic field energy density is at least comparable to the IR background, rather than escaping as gamma rays.  We consider two levels of pair cascading, which captures much of the energy output.  TeV photons, including those from IC, could escape if there is no other important background opacity than included here, implying a hard spectrum gamma-ray source with a sharp cutoff.

Fig.~\ref{stream} illustrates both the synchrotron and EBL cascade flux from residual escaping gamma rays (again using ELMAG \cite{Kachelriess2012} with extragalactic magnetic field to zero to maximize cascade flux).  There may be further extinction from the host galaxy on larger scales, although we do not attempt to include any such contribution (neither do we include X-ray absorption).  Also, while at low redshift the CMB can generally be counted as extragalactic for PeV gamma rays, at higher $z$ the attenuation length becomes much shorter ($n_{\rm CMB} \!\propto\! (1+z)^3$) and cause attenuation well within the host.  Generally there will also be model-dependent attenuation from within the acceleration region, which we discuss later without explicit inclusion here.

We see that the diffuse flux from cascades can be greatly diminished even only including this modest environmental component, if the source magnetic field strength is sufficient to override IC (which is somewhat Klein-Nishina suppressed for these energies).  For the Low-B case, the gamma rays are mostly pushed down to slightly lower energies and still contribute to the cascade background.  However, provided the magnetic field is at least comparable to the GC, where $B \!\gtrsim\! 8\,$mG was inferred by \citet{Eatough2013}, most of the pionic gamma-ray output is reprocessed via synchrotron that falls well below the cosmic X-ray \cite{Churazov2007} and MeV \cite{Weidenspointner2000} backgrounds (cf., \cite{Murase2015b}).  Including strong source evolution, described via the SFR \cite{Kistler2013b} in Fig.~\ref{stream}, further depresses the cascade flux below all the present data.  So, while details will depend on source properties, hard X-rays may offer an out and a path towards identifying the IceCube neutrino sources.

\section{Whence came the neutrinos?}
\label{whence}
Strictly speaking, we only know at this point that there is an abundance of $\sim\! 10 \!-\! 100$~TeV neutrinos and that the spectrum falls off at higher energies.  The {\it Fermi} IGB suggests that it is harder than $E^{-2}$ at lower energies if gamma rays can escape, perhaps favoring $p\gamma$ over $pp$ (see below though).  The situation at higher energies may be more complex (or less, or simply unrelated to what is going on at 10~TeV); however, we opt to follow the energy here as the $\gtrsim\,$PeV range is not as relevant in regard to total source energetics.

\subsection{Active Galactic Nuclei?}
The wide variety of AGN presents a great attraction to use the accretion energy to generate a relatively small number of very energetic protons in a compact volume (e.g., \cite{Margolis:1977wt,Berezinsky:1980mh,Kazanas:1985ud,Begelman1990,Stecker:1991vm,Nellen:1992dw,Halzen:1997hw,Rachen:1998fd,Muecke:2002bi,Atoyan:2002gu,AlvarezMuniz:2004uz,Kachelriess:2008qx,Stecker:2013fxa,Dimitrakoudis:2013tpa,Kimura:2014jba,Khiali2015}).  
At higher energies, comparisons of the $\sim\,$PeV IceCube flux to the UHECR spectrum in \cite{Kistler2014} suggested $\tau_{p\gamma} \!\lesssim\! 0.2$ (depending on assumptions about pion production).  At TeV energies, it is much more difficult to relate to the measured cosmic-ray flux since the Galactic flux is large, although more importantly such cosmic rays would not even be able to reach Earth if produced in regions where they are detained.  Although magnetic fields on cosmological scales are likely relevant, in an AGN there is always a supermassive black hole (SMBH).  Depending on the location of acceleration, escape may only mean being advected into the SMBH if entrained within the flow.

If the 10~TeV neutrinos arise from photopion production, the basic requirement is for a $\epsilon_\gamma \!\sim\! 0.2$~keV target background (assuming $z \!\sim\! 1$).  Then $\tau_{p \gamma} \!\propto\! L_{\rm keV}/R$ is $\sim\! 1$ for $L_{\rm keV} \!\sim\! 10^{43}\,{\rm erg/s}$ and $R \!\sim\! 10\,r_g$ for $r_g \!\simeq\! 1.5 \times 10^{12} (M/10^7\,M_\odot)$~cm.  This directly has implications for GeV $\gamma \gamma \!\rightarrow\! e^+ e^-$ extinction.  If, as in the GC a significant IR background results from stars, as evident in near-IR (e.g., \cite{Burtscher2015}), then the AGN mode is perhaps not as critical for $\gtrsim\! 10$~TeV $\gamma \gamma$ suppression, though one still needs to satisfy energetics requirements.

Taking a $z \!\approx\! 0$ AGN density of $n \!\sim\! 10^{-4}\, {\rm Mpc}^{-3}$ for $L_{\rm keV} \!\gtrsim\! 10^{42}\,$erg~s$^{-1}$  \cite{Hopkins:2006fq,Merloni2015}, apportioning the above emissivity implies $L_\nu \!\sim\! 10^{41} \!-\! 10^{42}\,$erg~s$^{-1}$.  This suggests that the responsible high-energy process should not be subtle, particularly if $\tau_{p \gamma} \!<\! 1$ so that there is even more energy in protons to account for, such emissivities could be comparable to a fraction of the total $z \!=\! 0$ AGN jet power density of $\sim\! 6 \!\times\! 10^{39}\,$erg~s$^{-1}\,{\rm Mpc}^{-3}$ \cite{Merloni2015}.

If all the relevant emission came from the same location, then it is simple to scale $\tau_{p \gamma}$ and $\tau_{\gamma \gamma}$.  However, the IR relevant for $\gtrsim\! 10$~TeV gamma rays can be much less compact than X-rays \cite{Antonucci1993,Urry1995,Netzer2015}, leading to a size regime investigated here (the GC CND itself has been compared to an AGN relic; \cite{Ponti2013}).  In AGN, driving energy comes from the accretion flow at the $\sim\! 10\,r_g$ scale, although the outer Bondi scale is $r_B \!=\! 2\,G\,M/v^2 \!\approx\! 1\,(M/10^7\,M_\odot)(v/300~{\rm km~s}^{-1})^{-2}$~pc, or $\sim\! 2 \!\times\! 10^6\, r_g$ for the nominal values.

Near-IR reverberation mapping and interferometry (e.g., \cite{Suganuma2006,Kishimoto2011}) have found dust ring radii scaling as the theoretical $r_d \!\approx\, 0.4 (L/10^{45}\,{\rm erg/s})^{1/2} (T_s/1500\,{\rm K})^{-2.6}\,$pc \cite{Nenkova2008}.  This is less so in the mid-IR, with radii $\sim\! 30$ times larger from interferometry and tending toward the $\sim$~pc scale \cite{Burtscher2013,Netzer2015}.  For comparison, in Fig.~\ref{ohigh} we include such a hotter $\gtrsim\! 1500$~K component (although to be more exact optically thick emission must be considered; e.g., \cite{Dermer1992,Dermer1993,Sikora2013}).

Even larger $\lesssim$~TeV extinction results if optical--UV is important, as from the hot thin disk in quasar models.  While this does not seem to be required at this point, greater attenuation could become relevant with better determination of IGB components.  This could also be relevant in suppressing synchrotron from muons/pions if important for neutrinos in the higher-energy PeV range \cite{Kistler2014,Winter:2013cla}.

\subsection{Low-Luminosity AGN and SMBH gaps?}
Such emission tends to be much less prominent in low-luminosity AGN (LLAGN) typically accounted for with advection-dominated accretion flow (ADAF) models (e.g., \cite{Yuan2014,Nemmen:2013mya}).  LLAGN are much more numerous than their brighter counterparts, with weaker cosmic evolution (e.g., \cite{Ho2008,Assef:2010ew}).  An ADAF can have large $\tau_{\gamma \gamma}$ for IR in inner accretion flow, as in the GC \cite{Kistler2015,Aharonian2005}.  However, the radiatively-inefficent clause of ADAFs presents difficulties for having sufficient $p \gamma$ targets if X-rays arise from the outer accretion flow (i.e., low compactness).  Sgr~A$^*$ is the rare instance where the X-ray emission can be resolved and a majority of the emission indeed appears to extend to the Bondi radius \cite{Wang2013}.

Acceleration may occur on a smaller scale than $p \gamma$ interactions, although there are not many smaller scales available than $10\, r_g$.  Recent observations of some highly-variable TeV sources have been ascribed to emitting regions relatively close to the SMBH (e.g., \cite{Acciari2009,Akiyama2015,Aleksic2014}).  The SMBH magnetosphere has been identified as a possible location where pulsar-like gaps can open on these small scales (e.g., \cite{Aleksic2014,Blandford:1977ds,Rieger:1999zv,Neronov:2007vy,Neronov:2007mh,Levinson:2010fc,Broderick:2015swa}).

We consider here a basic comparison of the ADAF conditions for gap formation, namely a local sub-Goldreich-Julian \cite{Goldreich:1969sb} charge density, $n_{\rm GJ}$, with those for neutrino production.
We take $L_\nu \!\sim\! 2 \!\times\! 10^{41}\,$erg~s$^{-1}$ (for $n \!\sim\! 10^{-4}\, {\rm Mpc}^{-3}$), a typical $E_p \!=\! 200$~TeV, and $\tau_{p\gamma} \!\sim\! 0.1$ for the above $L_{\rm AGN}$, since if the protons are back-lit a higher target $\epsilon_\gamma$ is needed due to the $p \gamma$ pion threshold, $\epsilon_\gamma \!\simeq\! m_p m_\pi/[E_p (1 - \cos{\theta_{p \gamma}})]$, generally decreasing $n_\gamma$.  For these, we estimate a proton current, $I_p \!\sim\! 6 \!\times\! 10^{39}\,{\rm s}^{-1}$, so that $n_p \!\simeq\! I_p/(4\pi R^2 c) \!\sim\! 0.8\, (1.5 \!\times\! 10^{14}{\rm cm}/R)^2\,{\rm cm}^{-3}$, for nominal $R \!\approx\! r_g$ for $M_{\rm BH} \!=\! 10^9 M_\odot$.

For $n_{\rm GJ} \!=\! \Omega B/(2 \pi e c) \!=\! 0.05\, \dot{m}^{1/2} M_9^{-3/2}\,{\rm cm}^{-3}$, with $\dot{m} \!\lesssim\! 2 \!\times\! 10^{-4}$ suggested to keep gaps open from $\gamma \gamma \!\rightarrow\! e^+ e^-$ \cite{Levinson:2010fc}, $n_p \!\gg\! n_{\rm GJ}$.  While large pair multiplicities $\mathcal{M} \!=\! n_{e^\pm}/n_{\rm GJ}$ can be produced in AGN via $\gamma \gamma \!\rightarrow\! e^+ e^-$ cascades downstream, this option is not available for protons.  Varying $M$ or $R$ tends to move both densities in the same direction (and we have not accounted for anisotropy of $I_p$ increasing $n_p$ or whether $\tau_{p\gamma}$ can actually be this large for such an $\dot{m}$) so would appear to be prima facie evidence disfavoring gaps as a primary TeV neutrino production site.

An alternative is acceleration in strong shocks within a collisionless ADAF environment.  GRMHD simulations with such hot accretion flows slightly tilted with respect to the SMBH spin axes have displayed strong shocks dissipating a substantial fraction ($\sim\!10$\%) of the accretion energy \cite{Generozov:2013mfa}.  At the high end of the ADAF regime with $\dot{m} \!\lesssim\! 10^{-2}$, X-ray luminosities may reach $\gtrsim\! 10^{42}\,$erg~s$^{-1}$ from the inner flow \cite{Yuan2014}.

For $L_p$ and $\tau_{p \gamma}$ both scaling with $\dot{m}$, one would expect the end of the population to dominate the neutrino output.  Needing a maximum $E_p \!\sim\! 200$~TeV is not as extreme as needed for UHECR and a tilt might be feasible if the flow is not MAD \cite{McKinney:2012wd} and may offer a way towards identifying the sources.  Similar to our above discussions, it is plausible that environmental effects are important, such as a MAD magnetic flux threshold for jet formation \cite{Igumenshchev2003,McKinney2012,Ghisellini2014}, so further study of effecting environments will be important.

In AGN, $pp$ might also be viable due to the more compact photon emission.  Even if the soft X-ray background is insufficient for a large $\tau_{p \gamma}$, $\tau_{\gamma \gamma}$ could still be relevant.  The result would be a photon spectrum only peaking out at the $\lesssim\,$GeV range, where the IGB is rising, though we defer details to elsewhere.

\subsection{Supernovae?}
Models obtaining large neutrino fluxes from starburst galaxies typically rely on $pp$ scattering, thus resulting in a spectrum extending to $\lesssim$~GeV (e.g., \cite{Bahcall1964,Loeb2006,Thompson:2006qd,Emig:2015dma,Ando2015}), with a flux that is perhaps now exceeded by the IceCube data for most parameter choices.  The gamma-ray spectrum would also be expected to similarly continue, with the most recent {\it Fermi} IGB determination \cite{Ackermann2015} and higher IceCube flux causing more tension than early indications \cite{Murase2013,Bechtol2015}.

Can this be alleviated?  The condition that we discussed earlier, $u_B$ at least as large as the relevant $u_{\rm IR}$ (i.e., accounting for Klein-Nishina suppression), may be satisfied in some starbursts (e.g., \cite{Thompson:2006is,Lacki:2010ue,McBride:2013mka,Yoast-Hull:2015iea,Beck2015}) if the IR compactness also allows sufficient $\tau_{\gamma \gamma}$ at $\gtrsim\!10$~TeV.  However, it does not appear that the starburst soft X-ray photon field would be opaque ($\tau_{\gamma \gamma} \ll 1$).  so while part of the flux might be eliminated, one would be left with the residual GeV--TeV emission.  One can also consider $p \gamma$ in these galaxies, however, the soft X-ray density needed appears insufficient relative to gas targets for $pp$.

Considering the local cosmic rate of core-collapse SNe, $\dot{n}_{\rm SN} \!\sim\! 10^{-4}\,{\rm Mpc}^{-3}{\rm yr}^{-1}$ (see collected data in \cite{Yuksel:2012zy,Horiuchi:2011zz}) and the SFR-normalized $\mathcal{E}_\nu \!\approx\! 1.5 \!\times\! 10^{37}\,$erg~s$^{-1}\,{\rm Mpc}^{-3}$ would imply $\sim\! 5 \!\times\! 10^{48}\,$erg of TeV neutrinos alone per SN (SNe~Ia have a subdominant rate and can occur anywhere; see \cite{Kistler:2011yk}).  If special conditions are required, such as low metallicity, these can be pushed out further in $z$, though only decreasing the per-event energetics demand by another $\sim\! 2$ even for GRB-like evolution \cite{Yuksel2007,Kistler2008,Kistler2013b}.

The spectrum below the 10~TeV break is also likely softer than $\alpha \!=\! -1$.  Taking $E_\nu^{-2}$ from the break back to $\sim\! 1$~GeV increases the neutrino energetics by $\sim\!4$.  Accounting for gamma-ray and secondary $e^\pm$ production adds another $\sim\!2$, so that the total reaches $\sim\! 4 \!\times\! 10^{49}\,$erg per SN (not including any primary $e^-$ acceleration), or $\sim\! 40$\% of the canonical $\sim\!10^{50}\,$erg of cosmic rays from {\it every} SN in the universe.

\section{Discussion and Conclusions}
\label{concl}
Aside from simply remaining below the present IGB, the gamma-ray flux from neutrino production will in the end need to fit into an overall picture that includes extrapolations from all sources and cosmogenic fluxes (e.g., \cite{Ajello2015,Murase2015b,Gelmini:2011kg,Kalashev:2013vba,Roulet:2012rv,Padovani:2015mba,Giacinti:2015pya}).  If the production mechanism is related to the flux of known sources, then there should be correlations.  Muons would be the preferred channel, although we found no clear coincidences between IceCube track events and {\it Fermi} 2FHL sources.  Further correlation \cite{Fang:2014uja,Petropoulou:2015upa,Adrian-Martinez:2015jlj} and spectral studies (e.g., \cite{Laha2013,Barger:2012mz,Vissani:2013iga,Anchordoqui:2013qsi,Chen:2013dza,Chen:2014gxa,Palomares-Ruiz:2015mka}) may clarify the relation to higher-energy neutrinos and whether any new physics \cite{Pakvasa:2012db,Barger:2013pla,Stecker:2014xja,Diaz:2013wia,Anchordoqui:2014hua,Learned:2014vya,Dutta:2015dka} is involved.

Otherwise, the associated gamma rays need to be suppressed.  This alone is not sufficient, as the energy they carry also needs to be shifted to some other band rather than just lower-energy gamma rays.  We have seen that plausible IR backgrounds surrounding a neutrino source can accomplish the first, while magnetic fields as strong as indicated in our Galactic Center can do the latter.  {\it NuSTAR} has recently measured a diffuse hard X-ray flux reaching to $\gtrsim\!40\,$keV pervading the GC \cite{Perez2015}, so perhaps such measurements can shed light on the IceCube sources.

While it is fairly straightforward to construct a source photopion neutrino spectrum harder than $E_\nu^{-2}$, to be as soft as $E_\nu^{-2.7}$ is more puzzling.  This could imply a spectrum of sources, in which case sources with a $E_p \!\gtrsim\! 100$~TeV cutoff dominate the emission.  The sources with higher cutoffs should have some combination of much lower number density and/or $\tau_{p \gamma}$, since the IceCube spectrum declines (for $\tau_{p \gamma}$ optically thick, the outgoing flux approaches the accelerated spectrum, e.g., $E^{-2}$).

If $\tau_{p \gamma}$ is already important at X-rays, one may wonder if acceleration to higher energies can proceed if lower-energy photons become more numerous and $p \gamma$ cooling becomes stronger with increasing proton energy.  Also, if $\tau_{p \gamma}$ becomes greater than unity, the source soon becomes opaque to $\tau_{n \gamma}$ so that contributions to the UHECR flux would be suppressed.

In total, while it may seem an obvious point that detecting more neutrinos is beneficial, these benefits need not be direct, e.g., by mapping count excesses.  Indirectly, extending the flux to lower energies and/or higher levels may start to break the simplest assumptions and force new insights into the nature of the sources of neutrinos and cosmic rays.\\

%
We thank Ranjan Laha, Greg Madejski, Matthew Wood, Neil Watkins, and Hasan Yuksel for useful discussions, the authors of ELMAG \cite{Kachelriess2012} for making their code available, and the INT Program INT-15-2a ``Neutrino Astrophysics and Fundamental Properties'' for hospitality during a portion of this project.
MDK acknowledges support provided by Department of Energy contract DE-AC02-76SF00515, and the KIPAC Kavli Fellowship made possible by The Kavli Foundation.


\end{document}